\documentstyle[aps,twocolumn,rotating,epsf,eqsecnum]{revtex}
\skip\footins 15.25pt plus 4pt minus 2pt
\def\footnoterule{\kern-5.25pt\hrule width.5in\kern3.6pt}
\renewcommand{\mathrm}[1]{{\rm #1}}

\begin{document}
\draft
\title{\large\bf
  Compound Poisson Statistics and Models of Clustering
of Radiation Induced DNA Double Strand Breaks.}

\vspace{24pt}
\author{ 
   E. {\sc Gudowska-Nowak},\ ${}^{1,2}$\\
 M. {\sc Kr\"amer},\ ${}^{2}$
G. {\sc Kraft}\ ${}^{2}$
and  G. {\sc Taucher-Scholz},\ ${}^{2}$}
   
\vspace{8pt}

\address{
${ }^{1}$ {\sl M. Smoluchowski Institute of Physics, Jagellonian 
University,}\\
{\sl ul. Reymonta 4,
30059 Krak\'ow, Poland}\\
${ }^{2}$ {\sl Biophysik, Gesellschaft f\"ur Schwerionenforschung, 
Planckstr. 1 , 64291 Darmstadt, Germany}}
    
\maketitle

\vspace{6pt}

\begin{abstract}

According to the experimental evidence damage induced by densely ionizing
radiation in mammalian cells is distributed along the DNA molecule in the form of clusters. The most critical constituent of DNA damage are double-strand  breaks
(DSBs) which are formed  when the breaks  occur in both DNA strands and are directly opposite or separated by only a few base pairs.  The paper discusses a model 
of clustered DSB formation  viewed in terms of compound Poisson process 
along with the
predictive assay of the formalism in application to experimental
data.\\
PACS numbers: 87.10.+e, 05.40.+j
\end{abstract}

\newcommand{\gm}{\gamma}
\newcommand{\ee}{\epsilon}
\renewcommand{\th}{\theta}
\newcommand{\Sg}{\Sigma}
\newcommand{\dl}{\delta}
\newcommand{\SSg}{\tilde{\Sigma}}
\newcommand{\eq}{\begin{equation}}
\newcommand{\eqx}{\end{equation}}
\newcommand{\eqn}{\begin{eqnarray}}
\newcommand{\eqnx}{\end{eqnarray}}
\newcommand{\ben}{\begin{eqnarray}}
\newcommand{\een}{\end{eqnarray}}
\newcommand{\f}[2]{\frac{#1}{#2}}
\newcommand{\ra}{\rangle}
\newcommand{\la}{\langle}
\newcommand{\bra}[1]{\la #1|}
\newcommand{\ket}[1]{| #1\ra}
\newcommand{\GG}{{\cal G}}
\renewcommand{\AA}{{\cal A}}
\newcommand{\GR}{G(\ee)}
\newcommand{\MM}{{\cal M}}
\newcommand{\BB}{{\cal B}}
\newcommand{\ZZ}{{\cal Z}}
\newcommand{\DD}{{\cal D}}
\newcommand{\HH}{{\cal H}}
\newcommand{\RR}{{\cal R}}
\newcommand{\arr}[4]{
\left(\begin{array}{cc}
#1&#2\\
#3&#4
\end{array}\right)
}
\newcommand{\arrd}[3]{
\left(\begin{array}{ccc}
#1&0&0\\
0&#2&0\\
0&0&#3
\end{array}\right)
}
\newcommand{\tr}{\mbox{\rm tr}\,}
\newcommand{\One}{\mbox{\bf 1}}
\newcommand{\pauli}{\sg_2}
\newcommand{\cor}[1]{<{#1}>}
\newcommand{\cf}{{\it cf.}}
\newcommand{\ie}{{\it i.e.}}
\newcommand{\br}[1]{\overline{#1}}
\newcommand{\phib}{\br{\phi}}
\newcommand{\psib}{\br{\psi}}
\newcommand{\zb}{\br{z}}
\newcommand{\qb}{\br{q}}
\newcommand{\lm}{\lambda}
\newcommand{\ksi}{\xi}

\newcommand{\Gb}{\br{G}}
\newcommand{\Vb}{\br{V}}
\newcommand{\Gm}{G_{q\br{q}}}
\newcommand{\Vm}{V_{q\br{q}}}

\newcommand{\ggd}[2]{\GG_{#1}\otimes\GG^T_{#2}\Gamma}

\section{Introduction}
In living cells subjected  to ionizing radiation many chemical reactions are induced leading to various biological effects such as mutations, cell lethality or neoplastic transformation \cite{HALL,KRAFT}. The most important target 
for radiation induced chemical transformation where these changes can be critical for cell survival is DNA distributed within the cell's nucleus. Nuclear DNA is organized in a hierarchy of structures which comprise the cellular chromatin. The latter is composed of DNA, histones and other structural proteins as well as
 polyamines. Organization of DNA within the chromatin varies with the cell type and changes as the cell progresses through the cell cycle. 
Ionizing radiation produces variety of damage to DNA including base alterations and single-  and
 double strand breaks (DSBs)
in the sugar-phosphate backbone of the molecule
\cite{HALL,WARD}. Single strand breaks (SSBs) are efficiently repaired with high fidelity and  probably contribute very little to the loss of function of living cells. On the other hand, DSBs are believed to be
the critical
lesions produced in chromosomes by radiation; interaction between DSBs can lead
to cell killing, mutation or carcinogenesis.
The purpose of theoretical modeling of radiation action \cite{TOBIAS}--\cite{SACHS} is to describe qualitatively
and  quantitatively the results of radiobiological effects at the molecular, chromosomal and cellular level. The basic consideration in such
an approach must be then descriptive analysis of breaks in DNA caused by charged particle tracks and
 by the chemical species produced.\\
Production of  DSBs  in intracellular DNA can be studied by use of
 the pulsed field gel electrophoresis (PFGE) \cite{ILIAKIS}
in which the gel electrophoresis is applied to elute high molecular weight DNA fragments from whole cellular DNA embedded in an organic gel (agarose).
Two main approaches of this technique are usually applied. One is the 
measurement of the fraction of DNA leaving the well in PFGE, i.e. the amount 
of DNA smaller than a certain cutoff size defined by the electrophoretic 
conditions. This method has proven to be very sensitive, allowing 
reproducible measurements at relatively low doses. The second approach is to 
describe fragment-size distributions obtained after irradiation as a 
function of dose, taking advantage of the property of PFGE to separate DNA 
molecules based on how quickly they reorient in a switching (pulsed) 
electrical field. The major goal of the experiments is to quantify number of 
induced DSBs based on changes in the amount of DNA or the average fragment 
size in response to dose. In both cases data obtained are related to average 
number of DSBs. To analyze the data, the formalism
describing random depolarization of polymers of finite size is usually adopted
\cite{MONTROLL,BLOCHER} giving  very well fits to experimental results with X-ray induced DNA fragmentation.
In contrast to the findings for sparsely ionizing irradiation (X and $\gamma$ rays) characterized by low average energy deposition per unit track length (linear energy transfer, LET$\approx$ 1 keV/$\mu$m), the  densely ionizing (high LET) particle track is  spatially  localized \cite{KRAFT,MICH}. In effect, 
multiplicity of  ionizations  within the track of heavy ions can produce clusters of DSBs on
packed chromatin \cite{NEWMAN}. The formation of clusters depends on chromatin geometry in the cell and radiation track structure.\\
DSBs multiplicity and location on chromosomes may determine the distribution of DNA fragments detected in PFGE experiments.
Modeling DNA fragment-size-distributions provides then a tool which allows to elucidate experimentally observed
frequencies of fragments. Even without detailed information on the geometry of chromatin, models of radiation
action on DNA can serve with some predictive information concerning measured DNA fragment-size-distribution.
The purpose of the present paper is to discuss a model which can be used in analysis of DNA
 fragment-size-
distribution  after heavy ion
  irradiation. The background of the model is the Poisson statistics of radiation events which lead to formation of clusters of DNA damage. The formation of breaks to DNA can be then described as the generalized or compound Poisson process for which the overall statistics of damage
is an outcome of the random sum of random variables (Section 2).
Biologically relevant distributions are further derived
and used (Section 3) in description of fragment size distribution in DNA after irradiation with
heavy ions. Practical use of the formalism is discussed by fitting the distributions to experimental data.
 
\section{Random sums of random variables and compound Poisson distributions}
Consider \cite{KAMPEN,PAP}  a sum $S_N$ of $N$ independent random variables $X$
\ben
S_N=\sum_{i=1}^N X_i
\een
where $N$ is a random variable with a probability generating function $g(s)$
\ben
g(s)=\sum^{\infty}_{i=0}g_i s^i
\een
and  $X_i$ are {\it i.i.d.} variables (independent and sampled from the same 
 distribution) whose 
  generating function $f(s)$ is
\ben
f(s)=\sum_{j=1}^{\infty}f_j s^j
\een
By use of the {\it Bayes rule} of conditional probabilities
the probability that $S_N$ takes value $j$ can be then written as
\ben
P(S_N=j)\equiv h_j= \sum^{\infty}_{n=0}P(S_N=j|N=n)P(N=n)
\label{zu}
\een
For fixed value of $n$ and by  using the statistical independence of $X_i$'s, the sum $S_N$ has a probability generating function $F(s)$
being a direct product of $f(s)$, {\it i.e.} $F(s)=f(s)^n=\sum^{\infty}_{j=0}F_js^j$
from which it follows that $P(S_N=j|N=n)=F_j$. The formula (\ref{zu})
leads then to
 the compound probability generating function of $S_N$ given by
\ben
h(s)=\sum^{\infty}_{j=0} h_js^j=\nonumber \\
=\sum^{\infty}_{j=0}\sum^{\infty}_{n=0}F_jg_ns^j=\nonumber \\
=\sum^{\infty}_{n=0}g_n{f(s)}^n\equiv g\{f(s)\}
\label{comp}
\een
Conditional expectations rules can be used to determine moments of a random sum. Given $E[N]=\nu$, $E[X_i]=\mu$, $Var[N]=\tau^2$ and $Var[X_i]=\sigma^2$,
the first and the second moment of the random sum $S_N$ are
\ben
E[S_N]=\mu\nu, \qquad Var[S_N]=\nu\sigma^2+\mu^2\tau^2
\label{mom1}
\een

The above compound distribution is  describing ``clustered statistics'' of
events grouped  in a number $N$ of clusters which itself has a distribution. As such, it is sometimes described in literature \cite{STUART} as ``mixture of distributions''. Out of many interesting biological applications of compound distributions \cite{GOEL}-\cite{KARLIN}, a special class constitute Poisson point processes which can be also analyzed in terms of random sums with Poisson distributed random events $N$.
It can be shown that a mixture of Poisson distributions resulting from using
any unimodal continuous function $f(\lambda)$ is  a unimodal discrete distribution. It is not so, however,
in case of unimodal discrete mixing. In particular, mixtures of Poisson-Poisson
or Poisson-binomial, known in literature as {\it Neyman distributions} \cite{NEYMAN} can exhibit
strongly multinomial character. 
By virtue of the above formalism  and by using  the formulae (\ref{comp})
, the generating function of the compound Poisson-Poisson distribution is:
\ben
g=\exp(-\lambda(1-f(s)))
\een
where the random variables $X_i$ 
 are distributed according to a Poisson law
\ben
f(s)=\exp(-\mu+\mu s)
\een
and the total $S_N$  is a random variable with a compound Poisson-Poisson ({\it Neyman type A}) distribution:
\ben
P(S_N=x)\equiv P(x;\mu,\lambda)=\sum_{N=0}^{\infty}\frac{(N\mu)^xe^{-N\mu}}{x!}\frac{\lambda^N e^{-\lambda}}{N!}
\label{neyA}
\een
for which the mean and variance are given by
\ben
E[x]=\mu \lambda, \qquad Var[x]=\lambda\mu(1+\mu)
\een

The resulting distribution can be  interpreted as a mixture of Poisson distribution with parameter $N\mu$ where $N$ (number of clusters) is itself Poisson  distributed with parameter $\lambda$.
Figures 1,2 present  function (\ref{neyA}) for two various sets of parameters $\lambda, \mu$.

The compound Poisson distribution (CPD) has a wide application in ecology, nuclear
chain reactions and queing theory \cite{TOBIAS,REID,KARLIN,NEYMAN}. It is sometimes known as the distribution
of a ``branching process''  and as such has been also
used to describe radiobiological effects of densely ionizing radiation in cells 
\cite{GOEL,ALBRIGHT,EWA,RITTER}. When a single heavy ion crosses a cell nucleus, it may produce DNA strand breaks and chromatin scissions wherever the ionizing track structure overlaps chromatin structure.
The multiple yield of such lesions depends on the radial distribution of deposited energy and on the microdistribution of DNA in the cell nucleus. The latter and the geometry of DNA coiling in the cell nucleus determine 
 number of crossings, the ``primary'' incidents leading to DSBs production.
By assuming for a given cell line, a ``typical'' average 
number $n$ of possible crossings per  particle traversal, 
 the distribution of the number of chromatin 
breaks $i$ can be modelled by a binomial law:
\ben
P(i|n)={n\choose i} p^i q^{(n-i)}
\een 
where
 $p$ is a probability that a chromatin break occurs at each particle crossing (and $q$ is the probability that it does not).
 The overall probability that $i$ lesions will be observed after $m$ independent particles traversed the nucleus is given by \cite{TOBIAS}

\ben
{\bf P}(i|\sigma,F,n)=
\sum^{\infty}_{m=1} \frac{(nm)!p^iq^{(nm-i)}(\sigma F)^m e^{-\sigma F}}
{i! (nm-i)! m!}
\label{jney}
\een
which is a compound {\it Neyman type B}
distribution obtained as a random Poisson sum 
of binomially distributed {\it i.i.d} variables. In the above presentation the average number of particles crossing the cell nucleus $\lambda$ is proportional to the absorbed energy (dose) and given by a product $\lambda=\sigma F$
of particle fluence $F$ and nuclear cross section $\sigma$. \\

\begin{figure}[ht]
\centerline{\epsfxsize=8truecm \epsfbox{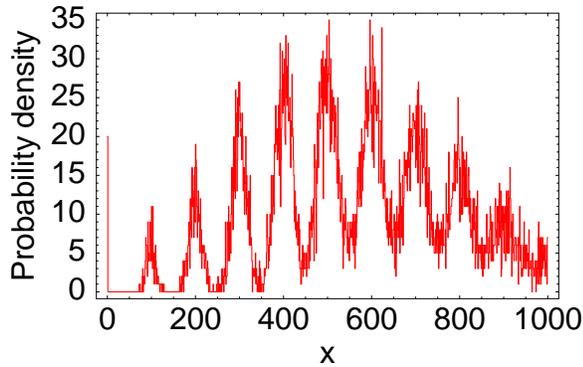}}
\vskip 0.5cm
\caption{Simulated probability density function for the Neyman-type A distribution (\ref{neyA}) with
 $\lambda=6, \mu=100$ for $N=10000$ points. Note the finite value at $x=0$ corresponding to $P(0;\mu, \lambda)$.}
\label{a1}
\end{figure}

Aggregation of observed cellular damage potentially leads to the phenomenon of ``overdispersion''-- that is,
the variance of the aggregate may be  larger than Poisson  variance yielding ``relative variance''
$Var_{rel}=Var[S_N]/E[S_N]$ larger than 1. Assuming thus the Poisson statistics of radiative events,
for any distribution of lesions per particle traversal, 
 the condition for overdispersion can be easily rephrased in terms of (\ref{mom1})
\ben
Var[X_i]/E[X_i]+E[X_i]<1
\een
If no repair process is involved in diminishing number of initially
produced lesions, the surviving fraction of cells can be estimated from  formula 
eq.(\ref{jney})  as a zero class of the initial
distribution, {\it i.e.} the proportion of cells with no breaks
\ben
{\bf P}_N(0|\sigma,F,n)=\sum^{\infty}_{m=1} \frac{(nm)!q^{nm}(\sigma F)^m e^{-\sigma F}}
{ (nm)! m!}=\nonumber \\
= \exp[-\sigma F (1-q^n)]
\een
which differs by a factor $(1-q^n)$ in the exponent from the surviving fraction for a Poisson distribution:
\ben
{\bf P}_P(0|\sigma,F,n)=\exp[-\sigma F]=\exp[-E[i]]
\een

\begin{figure}[ht]
\centerline{\epsfxsize=8truecm \epsfbox{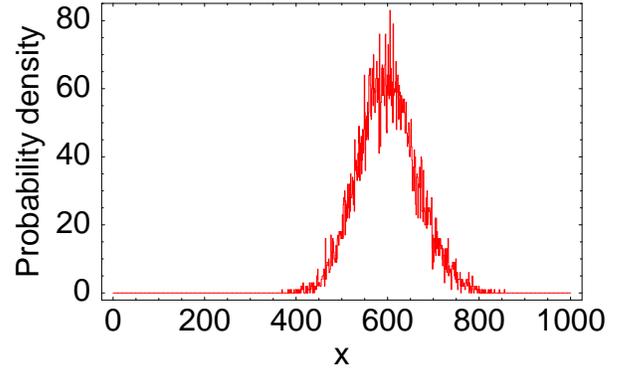}}
\vskip 0.5cm
\caption{Simulated probability density function for the Neyman-type A distribution (\ref{neyA}) with
 $\lambda=100, \mu=6$ for $N=10000$ points}
\label{a2}
\end{figure}

\section{DNA fragments distribution generated by irradiation: statistical model.}
DNA double stranded molecules in a size range from a few tenths of kilobase pairs to several megabase pairs can be evaluated by the PFGE technique. Randomly distributed DSBs are detected as smears
of DNA fragments.  The DNA mobility mass distribution may be transformed into a fragment length distribution using a calibration curve. It is obtained by relating migration distance of DNA within the gel to molecular length with the aid of size markers loaded on the same gel \cite{GISELA}. To interpret the experimental
material one needs to relate percentage of fragments in defined size ranges 
to number of induced DSBs. For that purpose several models have been derived,
mainly based on the description of random depolarization of polymers of finite
size \cite{MONTROLL,BLOCHER,RAD}. Although the models give satisfactory
prediction of size-frequency distribution of fragments after sparsely ionizing
radiation ({\it i.e} for X-rays and $\gamma$), they generally fail to describe
the data after densely ionizing radiation \cite{NEWMAN,GISELA}. The experiments with heavy ions demonstrate that after exposure to densely ionizing particles gives rise to substantially  overdispersed distribution of DNA fragments which indicates the occurrence of clusters of damage.
The following analysis presents a model which takes into account formation
of aggregates of lesions after heavy ion irradiation. \\   
Fragment distribution in PFGE studies is measured by use of  fluorescence technique or radioactive labeling with the result being the intensity distribution.
The generated signal is proportional to the relative intensity distribution of DNA fragments and  can be  expressed as
\ben
I(x)=xD(x)
\label{cont}
\een
with
\ben
D(x)=\sum_{j=0}^{\infty}D(x|j)P(j;\mu,\lambda)
\label{den}
\een
where $D(x|j)$ stands for the density of fragments of length $x$ provided $j$
DSBs occur on the chromosome of size $S$. Frequency distribution of the number of
DSBs is assumed here in the form of CPD (\ref{neyA}) with parameters $\mu$ and $\lambda$ representing average number of breaks produced by a single particle
traversal and  average number of particle traversals, respectively.  The ``broken-stick'' distribution \cite{FLORY,RAD} for $j$ breaks
on a chromosome of size $S$ yields a density of fragments of size $x$:
\ben
D(x|j)=\delta (x-S)+2j\frac{1}{S}(1-\frac{x}{S})^{j-1}+\nonumber \\
+  j(j-1)\frac{1}{S}(1-\frac{x}{S})^{j-1} 
\een
where the first two terms describe contributions from the edge fragments of the chromosome and the third term describes  contribution from the internal fragments of length $x<S$. The first term 
applies to the situation when $j=0$; the edge contribution can be understood by  observing
that the first and the $j+1$ fragment have the same probability of being size $x$.  Direct summation in formula (\ref{den})  leads to
\ben
D_N(x)= \exp(-\lambda(1-e^{-\mu})) \delta(x-S) + \nonumber \\
 +\frac{2\lambda\mu}{S}\exp(-\mu\frac{x}{S}+\lambda(e^{-\mu\frac{x}{S}}-1))+ \nonumber\\
 + e^{-\lambda}(1-\frac{x}{S})\frac{\mu^2 \lambda}{S}(1+ \lambda e^{-\mu\frac{x}{S}})
\exp(-\mu\frac{x}{S}+\lambda e^{-\mu\frac{x}{S}})
\label{ge}
\een
for Neyman distribution of number of breaks $j$ and to
\ben
D_P(x)=\Lambda\exp(-\Lambda\frac{x}{S})(2+\Lambda-\Lambda\frac{x}{S})
\een
for  a Poisson distribution with parameter $\Lambda$. 
Integration of $I(x)$ (eq.(3.1)) from 0 to some average (marker) size $X^*$ and division by $S$ yields the relative fraction of DNA content. For $\lambda>>1$ and
$\mu<<1$, the Neyman-type A distribution converges to a simple Poisson.
In such a case, simplified expression (\ref{ge}) leads to results known
in literature as ``Bl\"ocher formalism'' \cite{MONTROLL,BLOCHER,RAD} which describes
well the DNA content in probes irradiated with X-- and $\gamma$--rays.\\
\begin{figure}[ht]
\centerline{\epsfxsize=8truecm \epsfbox{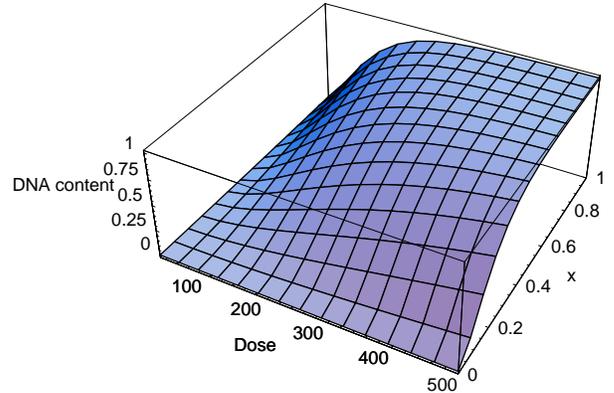}}
\caption{Distribution of DNA content (integrated eq.(\ref{cont}))
as a function of the dose and fragment size for
 $S=245 Mbp, \mu=5$. The fragments length is in Mbp units.}
\label{obrazek1}
\end{figure}
 Figure 3  presents predicted dose-response curves for the model. The amount of DNA content is
shown in function of dose and fragment size. In calculation, the parameter $S=245$ mega base pairs
has been used which is the mean chromosome size for Chinese hamster cells, the cell line for which
experimental data are displayed in Figure 4.\\
The increase in multiplicity of DSBs produced per one traversal of a particle leads to pronounced
increase in production of shorter fragments which is illustrated in the shift of the peak intensity towards smaller $x$ values.    
\begin{figure}
\centerline{\epsfxsize=8truecm \epsfbox{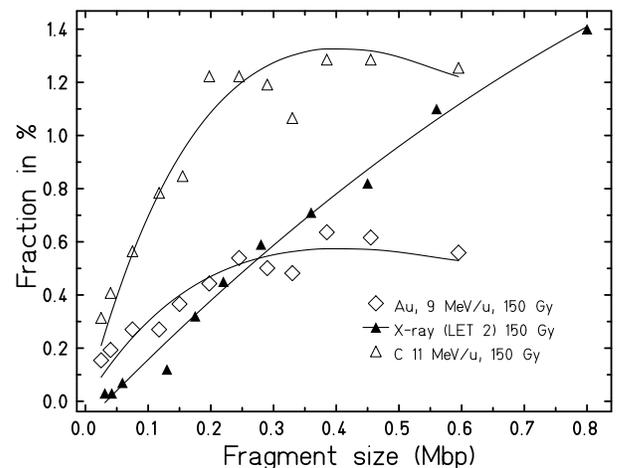}}
\caption{Fraction  of DNA content observed experimentally within the range of sizes 0.1-1.0 Mbp. 
Data show higher probability of producing short fragments after irradiation with particles
than for sparsely ionizing radiation at comparative dose. Lines represent the
 best fit to eq.(3.1) by use of $D_N(x)$ function for heavy ions (Au: $\lambda= 3\times 10^{-3}, \mu=6\times 10^2$ ; C: $\lambda= 6\times 10^{-3}, \mu=6\times 10^2$)  and $D_P(x)$ for
X-rays ($\Lambda =0.85$).}
\label{obrazek2}
\end{figure}

\section{Spatial Clustering of Breaks and Non-Poisson statistics.}

Clustering of breakage events can be viewed as the process leading to
 non-exponential ``spacing'' between subsequent events,  similar to the standard analysis of level repulsion in spectra 
of polyatomic molecules and complex nuclei. For a random sequence, the probability that a DSB will be in the
 infinitesimal interval 
\ben
(X+x, X+x+dx)
\een
proportional to $dx$ is independent of whether or not there is a break at $X$. This result can be easily changed by using the concept of breaks ``repulsion'. Given a break at $X$, let $P(x)dx$ be the probability that the next break  ($x\ge 0$) be found in the interval $(X+x, X+x+dx)$. We then have for the nearest-neighbour spacing  distribution of breaks the following formula:
\ben
P(x)dx=Prob(1\in dx|0 \in x)Prob(0 \in x)
\een 
where  $Prob(n\in dx|m\in x)$ is the conditional probability that the infinitesimal interval of length $dx$ contains $n$ breaks wheras that of length $x$ contains $m$ of those.
The first term on the right-hand side of the above equation  is $dx$ times a function of $x$ which we denote by $r(x)$, depending explicitly on the choices 1 and 0 of the discrete variables $n$ and $m$. The second term is given by the probability that the spacing is larger than $x$:
\ben
\int^{\infty}_x P(y)dy
\een 
 Accordingly, one  obtains
\ben
P(x)=r(x)\int^{\infty}_x P(y)dy,
\een
whose solution can be easily found to be
\ben
P(x)=C r(x)\exp(-\int^x r(y)dy)
\een
where $C$ is a constant. The Poisson law, which reflects lack of correlation between breaks,  follows if one takes $r(x)=\lambda$,
where $\lambda^{-1}$ is the mean spacing between DSBs. If choosing on the other hand
\ben
r(x)=\lambda x^{\lambda-1}
\een
{\it i.e.} by assuming clustering of points (DSBs) along a line, one ends up with the {\it Weibull density}.
The constants $C$ and $\lambda$  can then be determined from appropriate  conditions, {\it e.g.}
\ben
\int P(x)dx=1,
\een
and
\ben
\int x P(x)dx=\lambda^{-1}
\een
One then finds that
\ben
P(x)=\lambda e^{-\lambda x}
\label{exp}
\een
for the Poisson distribution and
\ben
P(x)=\lambda x^{ \lambda-1} \exp(- x^{\lambda})
\label{wig}
\een
for the Weibull analogue. Note that the above density
can be derived as a generalization of the law eq.(\ref{exp}): the Weibull density can be obtained as the density of random variable $y=x^{1/\lambda}$ with $x$ being an exponential random variable. For $\lambda\ge 1$, the Weibull distribution is unimodal with a maximum at point $x_m=(1-\lambda^{-1})^{\lambda^{-1}}$.
In this one easily recognizes for $\lambda=2$ the spacing distribution of the {\it Wigner law}. The latter  displays ``repulsion'' of spacing, since $P(0)=0$,
in contrast to the Poisson case which gives maximum at $x=0$. Fractional exponent
$\lambda< 1$ describes, on the other hand, enhanced frequency of short spacings
which, in fact, matches better experimental data for heavy ions({\it cf.} Figure 4).
The above analysis brings also similarities with random walks \cite{REICHEL,NELSON} where symmetry breaking transition manifests
itself as a change in the spectral spacing statistics of decay rates.  
In such cases, the statistics of events of interest deviates, as a counting process, from the regularity
of Poisson process, for which the subsequent event arrivals are spaced with a constant mean $\lambda^{-1}$.
The clustered statistics of breakage can be thus viewed as a (fractal) random walk or a cumulative distribution of a random sum of random variables eq.(2.1). The problem of characterizing the limit distribution for such cases with underlying ``broad'' distributions $g(x)$ of $X_i$ has been studied extensively in mathematical literature \cite{ZOLOTAREV} and has been solved with classification of the possible limit distributions provided
that requirement of ``stability'' is fulfilled under convolution. Following the definition, the distribution
$g(x)$ is  stable, if for any $N$ there exist constants $c_N>0$ and $d_N$ such that
$S_N$ has the same density as the variable $y=c_N X_i+d_N$. The stability condition
can be rephrased in terms of the canonical representation given by a form of the characteristic function ({\it i.e.} the Fourier transform $g(k)$)  of stable distributions \cite{KOLMOGOROV,ZOLOTAREV}
\ben
\ln g(k)=i\gamma k -C |k|^\lambda[1-i\omega \beta sign(k)]
\een
where $\gamma$ is real, $0\le \lambda\le 2$, $\omega$ is real and $|\omega|\le tan(\pi\lambda/2)|$. The cases relevant for biological modelling are covered by $1\le \lambda\le2$ (stable distributions have no variance if $\lambda<2$ and no mean if $\lambda<1)$.
In particular, positivity of steps in the random walk modelled by eq.(2.1)
allows for $g(k)=\exp[-C|k|^\lambda]$ which gives asymptotically $g(x)\approx x^{-\lambda-1}$. Probability distribution  that $x\ge z$ satisfies then $f(z)\approx z^{-\lambda}$
for $z\rightarrow\infty$. The resulting distribution  is ``self-similar'' in the sense
that rescaling $z$ to $Az$ and $f(z)$ to $A^{-\lambda}f(z)$ does not change the power law distribution. In other words, the number of realizations larger than $Az$ is $A^{-\lambda}$ times the number of realizations larger than $z$. The power-law probability distribution function describes then the same proportion of shorter and larger fragments whatever size is discussed within the power law range. For $\lambda=1/2, C=1, \omega=1$  the form of  L\`evy-Smirnov law is recovered
\ben
g(x)=(2\pi)^{-1/2}x^{-3/2}e^{-\frac{1}{2x}}
\label{smir}
\een 
The probability density eq.(\ref{smir}) has a simple
 interpretation as the limiting law of return times to the origin for a one-dimensional symmetrical random walk and as such has  been also used
to describe the fragment size distribution of a one dimensional polymer \cite{WEISS,PONOMAREV}. 
In the problems related to polymer fragmentation induced by irradiation, the approach based on a random walk with fluctuating number of steps (or, equivalently, on a point proceses model with a clustered statistics of waiting times) is a legitimate one as it can comprise 
the natural randomness of primary events ({\it i.e.} particle hits of biological target)
and secondary induction of multiple (clustered) lesions. Further investigations in this field should lead to better understanding of possible emergence of  power-law distributions 
of larger fragments on kbp and Mbp scales.

\section{Conclusions}
An existing  substantial evidence demonstrates that exposure to densely ionizing charged particles gives rise to overdispersed distribution of chromatin breaks and DNA fragments which is indicative
of clustered damage occuring in irradiated cells. The clustering process can be expressed for any particular class
of events such as ionizations or  radical species formation and is a consequence of energy localization in 
the radiation track. Chromosomal aberrations expressed  in irradiated cells are formed in process of misrejoining
of fragments which result from production of double-strand breaks in DNA. The location of double-strand breaks along chromosomes determines DNA fragment-size distribution which can be observed experimentally.
The task of stochastic modeling is then to relate parameters of such distributions to relevant 
quantities describing number of induced DSBs. Application of  the formalism of clustered breakage  
 offers thus a tool in   evaluation of the radiation respone of DNA fragment-size distribution and  assessment of radiation induced biological damage. \\ \\ 

\noindent{\bf Acknowledgements}.\\
\noindent E.G-N acknowledges partial support by KBN grant 2PO3 98 14 and by KBN--British Council
collaboration grant C51.\\ \\ 
\vskip 0.5truecm

\end{document}